\documentclass[a4paper,12pt, oneside]{article}
\usepackage{graphicx}
\usepackage{dcolumn}
\usepackage{bm}

\begin{document}
\baselineskip 24pt

\begin{center}
{\Large X-ray magnetic circular dichroism characterization of \\GaN/Ga$_{1-x}$Mn$_x$N 
digital ferromagnetic heterostructure}
\vspace{1cm}

{J. I. Hwang, M. Kobayashi, G. S. Song and A. Fujimori}\\
\it{Department of Complexity Science and Engineering and Department of Physics, University of Tokyo, Tokyo, 113-0033, Japan}
\vspace{0.5cm}

\rm{A. Tanaka}\\
\it{Department of Quantum Matter, ADSM, Hiroshima University, Hiroshima 739-8530, Japan}
\vspace{0.5cm}

\rm{Z. S. Yang, H. J. Lin, D. J. Huang and C. T. Chen}\\
\it{National Synchrotron Radiation Research Center, Hsinchu 30076, Taiwan}
\vspace{0.5cm}

\rm{H. C. Jeon and T. W. Kang}\\
\it{Quantum-Functional Semiconductor Research Center and Department of Physics, Dongguk University, Seoul 100-715, South Korea}
\vspace{0.5cm}

\end{center}


\newpage
\begin{center}
\section*{Abstract}
\end{center}
We have investigated the magnetic properties of a GaN/Ga$_{1-x}$Mn$_x$N ($x$ = 0.1) digital ferromagnetic heterostructure (DFH) showing ferromagnetic behavior using soft x-ray absorption spectroscopy (XAS) and x-ray magnetic circular dichroism (XMCD). 
The Mn $L_{2,3}$-edge XAS spectra were similar to those of Ga$_{1-x}$Mn$_x$N random alloy thin films, 
 indicating a substitutional doping of high concentration Mn into GaN. 
From the XMCD measurements, it was revealed that paramagnetic and ferromagnetic Mn atoms coexisted 
in the Ga$_{1-x}$Mn$_x$N digital layers. 
The ferromagnetic moment per Mn atom estimated from XMCD agreed well with that estimated from SQUID measurements. From these results, we conclude that the ferromagnetic behavior of the GaN/Ga$_{1-x}$Mn$_x$N DFH sample arises only from substitutional Mn$^{2+}$ ions in the Ga$_{1-x}$Mn$_x$N digital layers and not from ferromagnetic precipitates. Subtle differences were also found from the XMCD spectra between the electronic states of the ferromagnetic and paramagnetic Mn$^{2+}$ ions.


\newpage

%
%
%
%
Diluted magnetic semiconductors (DMS) have recently been studied extensively because of their potential applications for spintronics. For potential applications, it is indispensable to synthesize DMS with Curie temperatures ($T_C$'s) above room temperature. GaN-based DMS have attracted particular interest because the GaN host has been considered as a promising host material for 
DMS having high $T_C$ based on theoretical predictions \cite{Dietl, Mah}. 
Although Ga$_{1-x}$Mn$_x$N random alloy thin films were successfully grown by molecular-beam epitaxy (MBE) on sapphire substrates \cite{gmn1, gmn2, gmn3, gmn4}, a high concentration of Mn atoms cannot be incorporated into the GaN host because nitrogen defects are inevitably created in the crystals. 
Furthermore, it has been recognized that ferromagnetic precipitates such as Ga$_x$Mn$_y$ having $T_C$ exceeding room temperature can be formed in Ga$_{1-x}$Mn$_x$N \cite{4GaMn}. These facts have made the realization of high $T_C$ DMS difficult and search for the origin of the ferromagnetism complicated. 
On the other hand, a recent Monte Carlo simulation study \cite{Monte} has predicted that the $T_C$ of Ga$_{1-x}$Mn$_x$N random alloy remains low 
because the magnetic interaction between Mn ions in the GaN host is short ranged and consequently long range ferromagnetic ordering is not realized for low Mn concentrations. Therefore, doping high concentration Mn into the GaN host without degrading its quality has been desired to realize Mn-doped GaN having a high $T_C$. 
It has been reported that high quality Ga$_{1-x}$Mn$_x$N thin films grown by MBE on wurtzite GaN substrates showed ferromagnetism but with low $T_C$ \cite{5SonodaAPL, 5SariPRB}. 
It has been demonstrated that the ferromagnetism is enhanced for samples in which Mn$^{2+}$ and Mn$^{3+}$ coexist, indicating the possibility of high temperature ferromagnetism in this system \cite{5SonodaAPL}. Thus, recent studies have focused on the reduction of N defect concentration and the increase of Mn concentration in Ga$_{1-x}$Mn$_x$N \cite{quality}.

An interesting and promising route to realize high $T_C$'s is to synthesize DMS embedded in two-dimensional (2D) heterostructures by digital doping. The heterostructures consist of non-magnetic semiconductor layers and highly doped thin magnetic semiconductor layers, and are called digital ferromagnetic heterostructure (DFH). This allows the doping of high concentration transition-metal ions into spatially localized regions in the semiconductor host without degrading its quality. Nazmul $et\ al.$ \cite{4deltaGMA} reported that Mn delta-doped GaAs indicated ferromagnetism up to $T_C$ = 250 K. 
Jeon $et\ al.$ \cite{4DFH} adopted the DFH technique to the growth of Ga$_{1-x}$Mn$_x$N to enhance the local Mn concentration and reported that a considerable enhancement of magnetization and $T_C$ was achieved compared with those of Ga$_{1-x}$Mn$_x$N random alloy. 

In the present study, we have performed high energy spectroscopic studies of Ga$_{1-x}$Mn$_x$N DFH using 
x-ray absorption spectroscopy (XAS) and x-ray magnetic circular dichroism (XMCD) to investigate the electronic structure and the magnetic property of the GaN/Ga$_{1-x}$Mn$_x$N DFH. 
XMCD is an element selective technique and is a powerful tool to investigate the microscopic origin of ferromagnetism in solids. Subsequently, theoretical analysis based on configuration-interaction theory was adopted to understand the local electronic structure of GaN/Ga$_{1-x}$Mn$_x$N DFH.

\begin{figure*}[tttt]
\begin{center}
\includegraphics[height=4.4cm]{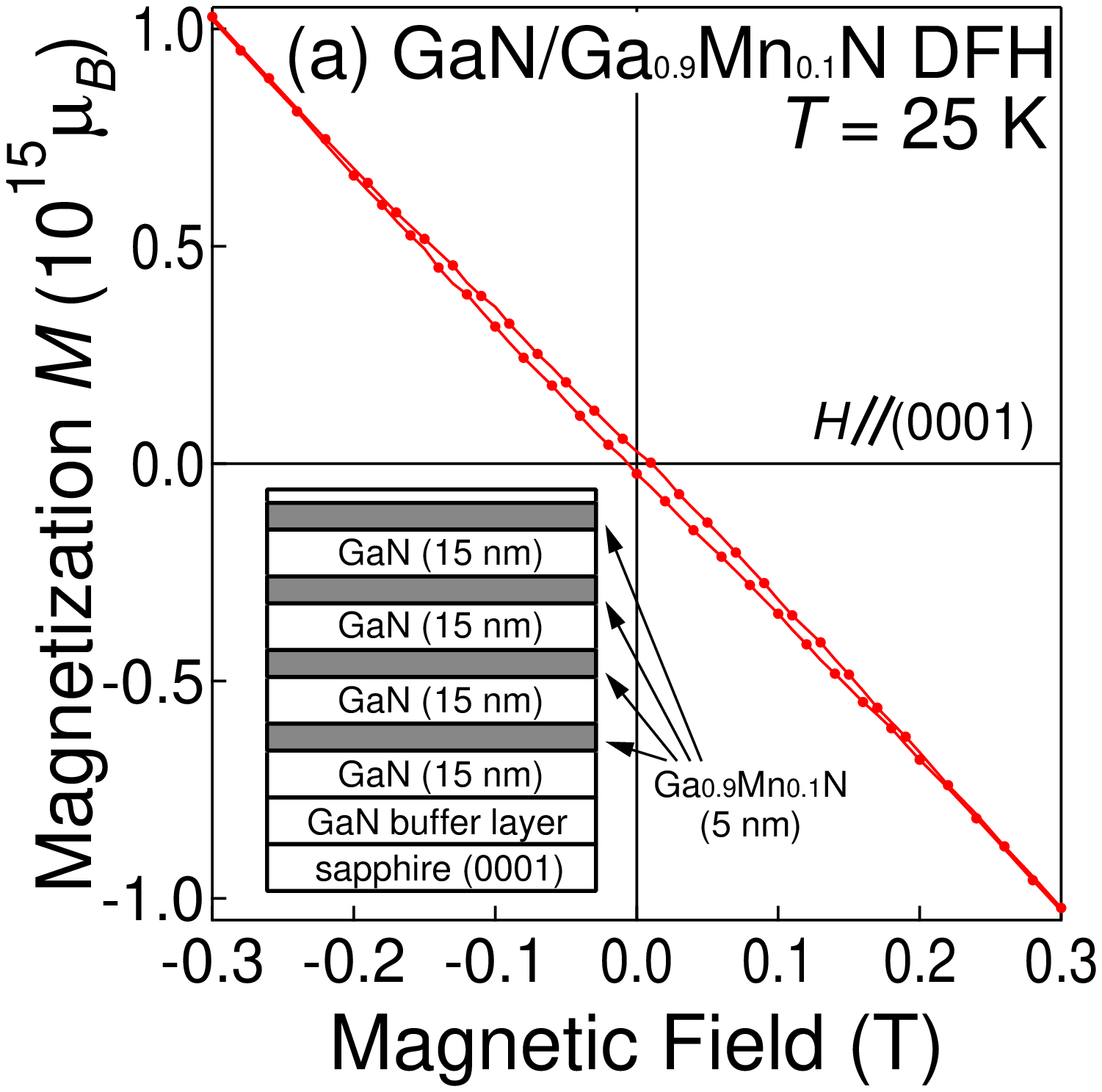}
\includegraphics[height=4.4cm]{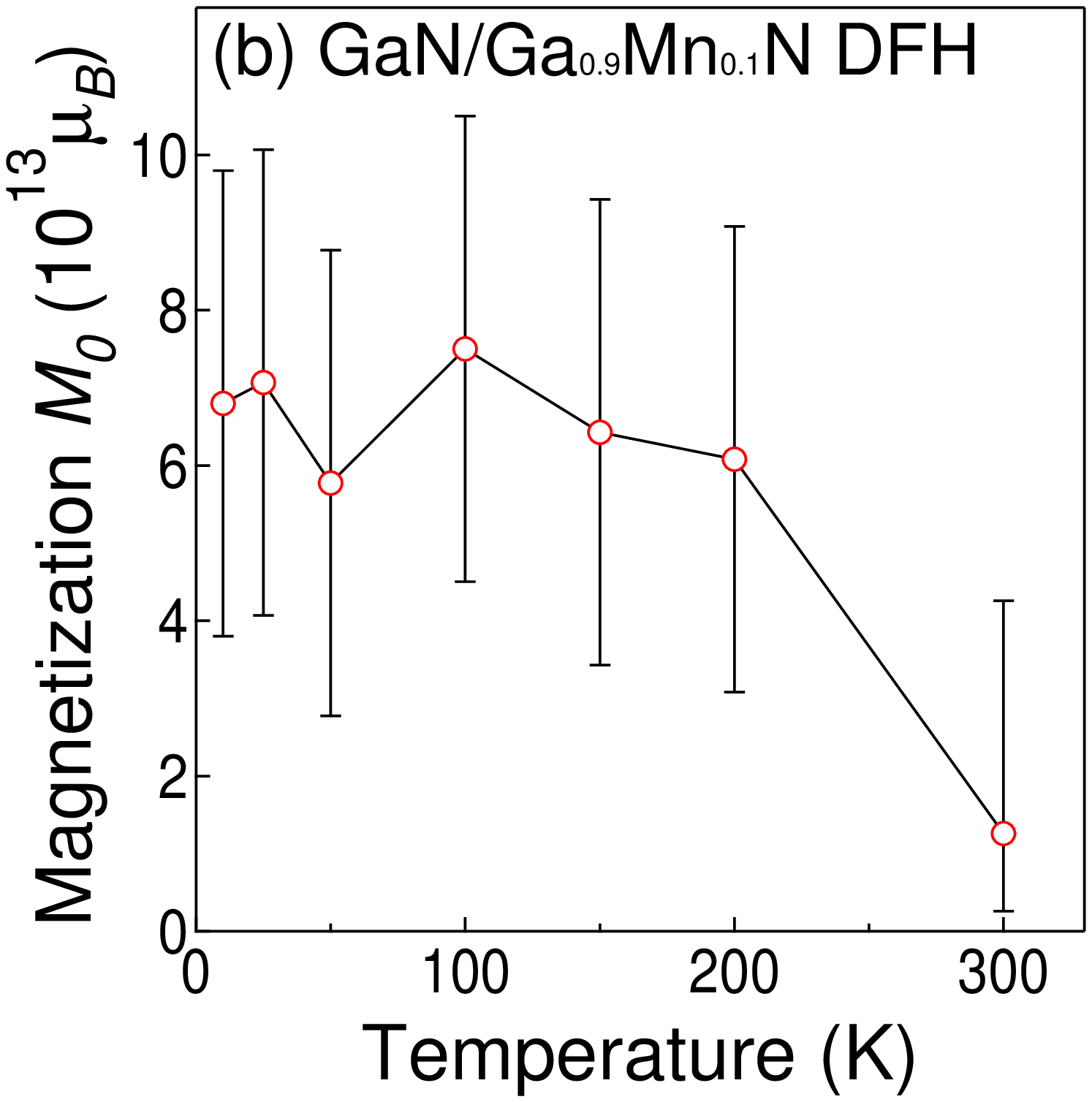}\\
\vspace{0.5cm}
\includegraphics[height=4.1cm]{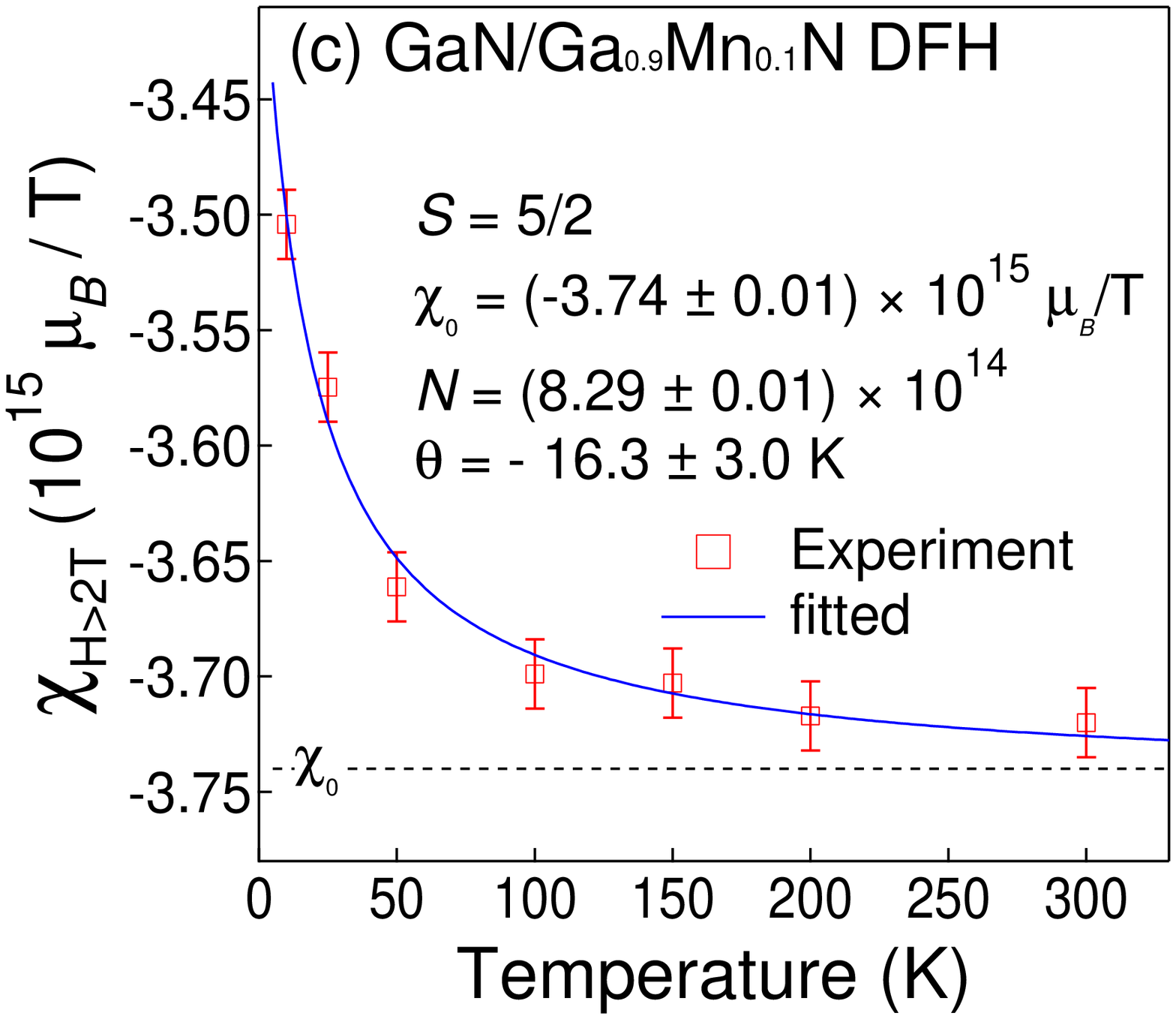}
\includegraphics[height=4.1cm]{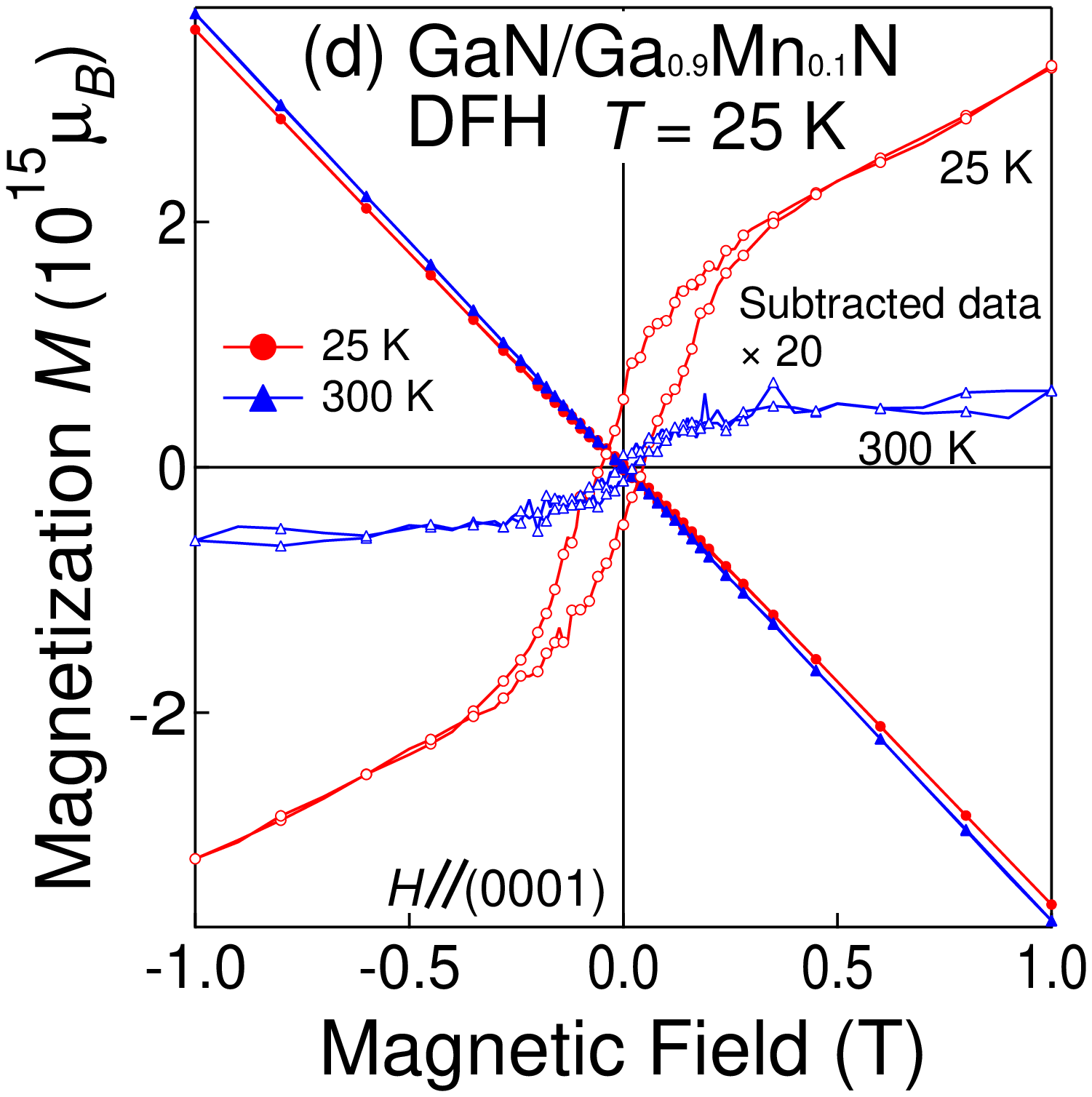}
\caption{(Color online) Magnetization measurements of the GaN/Ga$_{1-x}$Mn$_x$N DFH using a SQUID magnetometer. (a) Magnetization curve measured at $T$ = 25 K. The inset shows the structure of the GaN/Ga$_{1-x}$Mn$_x$N DFH sample. (b) Temperature dependence of the saturation magnetization.  (c) Temperature dependence of the high-field magnetic susceptibility. 
(d) Magnetization curve at 25 K (open circle) and 300 K (open triangle) attributed to the Ga$_{1-x}$Mn$_x$N layers in the DFH.}
\label{squid}
\end{center}
\end{figure*}

The GaN/Ga$_{1-x}$Mn$_x$N DFH sample used in this study was grown on a sapphire substrate by the RF-plasma-assisted MBE \cite{4DFH}. A multiple heterostructure of GaN/Ga$_{1-x}$Mn$_x$N of 4 periods was grown on the sapphire substrate with a GaN buffer layer at the substrate temperature of 850 $^{\circ}$C. The sample structure is shown in the inset of Fig. \ref{squid} (a). The thickness of each Ga$_{1-x}$Mn$_x$N layer and GaN layer was 5 and 15 nm, respectively. On top of it, a thin GaN capping layer ($<$1 nm) was deposited to prevent oxidations of the top Ga$_{1-x}$Mn$_x$N layer.


XAS and XMCD measurements were performed at the Dragon beam line BL11A of Taiwan Light Source (TLS). Absorption spectra were measured by the total electron yield method with the energy resolution $E$/$\Delta$$E$ better than 10000 and the circular polarization of 83 \%. In the XMCD measurements, the circular polarization of the incident photons was fixed and the direction of the applied magnetic field was changed. The XAS and XMCD measurements were made at temperature of 25 K in an ultra-high vacuum below 1$\times$10$^{-10}$ Torr. Magnetization measurements were made using a superconducting quantum interference device (SQUID, MPMS, Quantum Design Co., Ltd.) at the Institute for Solid State Physics (ISSP), University of Tokyo.

Figure \ref{squid} (a) shows the magnetization curves of the GaN/Ga$_{1-x}$Mn$_x$N DFH measured at $T$ = 25 K using the SQUID magnetometer. The magnetic field was applied perpendicular to the sample surface [the (0001) direction of the wurtzite structure]. One can see a hysteresis loop superimposed on a negative slope, indicating that the present DFH sample showed ferromagnetism at low temperature. The linear response of the magnetization $M$ in the high magnetic field region ($H>2$ T) was fitted to the formula $M = \chi_{H>2{\rm T}} H + M_0$, where $\chi_{H>2{\rm T}}$ and $M_0$ are high-field magnetic susceptibility and the saturation magnetization, respectively. Figure \ref{squid} (b) shows the temperature dependence of the saturation magnetization $M_0$ thus obtained. One can see that the saturation magnetization rapidly decreased above 200 K forward zero at 300 K, implying that the $T_C$ of the present sample was $\le$ 300 K. 

Figure \ref{squid} (c) shows the temperature dependence of the high-field magnetic susceptibility $\chi_{H>2{\rm T}}$. The data were fitted to the Curie-Weiss law plus a temperature-independent constant $\chi_0$: $\chi_{H>2{\rm T}} = {\partial M}/{\partial H_{H>2{\rm T}}} = NC/(T-\Theta) + \chi_0$. Here, $N$ is number of the magnetic Mn ions, $g$ is the $g$ factor, $C = (g\mu_B)^2S(S+1)/3k_B$ is the Curie constant and $\theta$ is the Weiss temperature. Assuming $S=5/2$ (Mn$^{2+}$), a good fit was obtained with $N = 8.29$ ($\pm$ 0.01) $\times 10^{14}$, $\Theta = - 16.3 \pm 3.0$ K and $\chi_0 = - 3.74$ ($\pm$ 0.01) $\times 10^{15}\mu_{\rm B}/{\rm T}$, indicating that the temperature dependence of the susceptibility was caused by local magnetic moments in the paramagnetic state. The negative Weiss temperature indicates a weak antiferromagnetic interaction between the local moments. The amount of magnetic ions contributing to the paramagnetism plus ferromagnetism was $\sim$ 58 \% of the total Mn ions doped into the sample. This indicates that $\sim$ 40 \% of the Mn ions are magnetically inactive, probably due to strong antiferromagnetic coupling between the Mn ions. The negative constant $\chi_0$ was due to the diamagnetic contribution from the GaN layers in the DFH and the sapphire substrate. 
Subtracting the diamagnetic component from the raw data, one can obtain the magnetization curve attributed to Ga$_{1-x}$Mn$_x$N layers in the DFH as shown in Fig. \ref{squid} (d). The magnetization curve thus obtained showed a clear hysteresis loop at 25 K while the hysteresis was not observed at 300 K. The magnetization at 25 K was not saturated even at $H$ = 1.0 T due to the paramagnetic contribution while the hysteresis loop was closed at $H$ = 0.4 T.

\begin{figure}[t]
\begin{center}
\includegraphics[height=4.9cm]{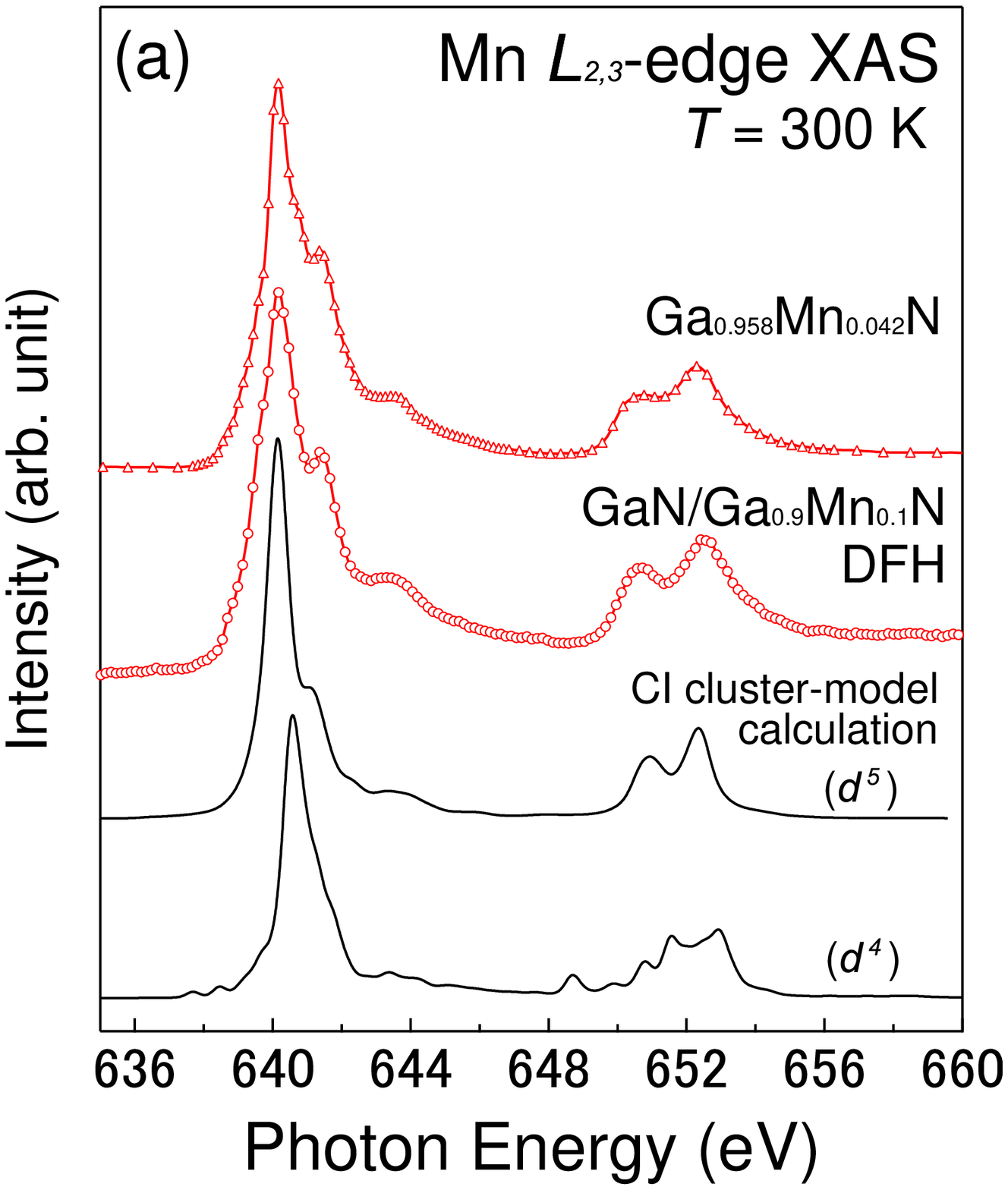}
\includegraphics[height=4.9cm]{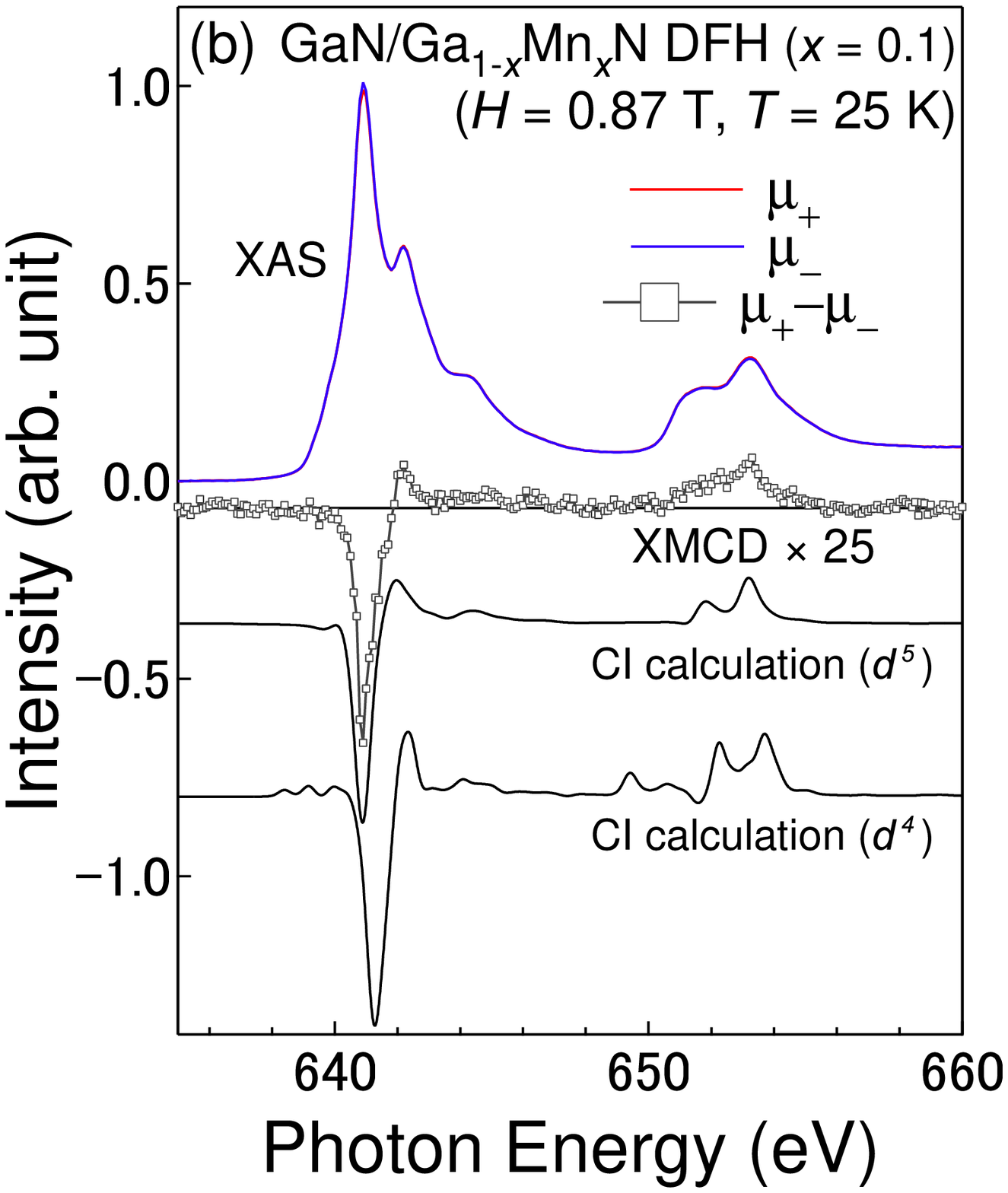}
\caption{(Color online) X-ray absorption (XAS) spectra and x-ray magnetic circular dichroism (XMCD) at the Mn $L_{2,3}$ edge of GaN/Ga$_{1-x}$Mn$_x$N DFH. (a) X-ray absorption spectra of GaN/Ga$_{1-x}$Mn$_x$N DFH (open circles) compared with that of the Ga$_{1-x}$Mn$_x$N ($x$ = 0.042) random alloy (open triangles) \cite{hwang} and the CI cluster-model calculations for $d^4$ and $d^5$ ground states in a crystal field of $T_d$ symmetry (solid line) using the electronic-structure parameters $\Delta = 4.0$ eV, $U = 5.0$ eV and $(pd\sigma) = -1.3$ eV. (b) XAS and corresponding XMCD spectra recorded at $T$ = 25 K and $H$ = 0.87 T.}
\label{mcd}
\end{center}
\end{figure}

To investigate the microscopic nature of the ferromagnetic and paramagnetic components, we have performed x-ray magnetic circular dichroism (XMCD) measurements at the Mn $L_{2,3}$ edge. Figure \ref{mcd} (a) shows the Mn $L_{2,3}$-edge XAS spectra of the GaN/Ga$_{1-x}$Mn$_x$N DFH compared with that of the Ga$_{1-x}$Mn$_x$N random alloy ($x$ = 0.042) \cite{hwang} and the configuration-interaction (CI) cluster-model calculations. The parameters used in the CI cluster-model calculations are the charge-transfer energy from the ligand $p$ orbitals to the transition metal $d$ orbitals ($\Delta$), the on-site Coulomb energy between two 3$d$ electrons ($U$) and the $p$-$d$ hybridization strength 
defined by a Slater-Koster parameters ($pd\sigma$) \cite{cicalc}, 
The rich structures in the observed spectrum are also close to those of the random alloy and can be explained by the typical multiplet structure of the localized $d^5$ (Mn$^{2+}$) configuration, and not of the $d^4$ (Mn$^{3+}$) configuration.
The CI cluster-model calculation assuming the $d^5$ ground state indeed well reproduced the multiplet structure of the experimental spectra using $\Delta = 4.0 \pm 0.5$ eV, $U = 5.0 \pm 0.5$ eV and $(pd\sigma) = -1.3 \pm 0.2$ eV. This lead us to conclude that the Mn ions were in the high-spin Mn$^{2+}$ ($d^5$) state in a crystal field of $T_d$ symmetry. This result indicates 
that a high concentration Mn ions were substitutionally doped at the Ga site without segregations of impurity phases.

In Fig. \ref{mcd} (b), we compare the XMCD spectrum of the GaN/Ga$_{1-x}$Mn$_x$N DFH with the CI cluster-model calculations for $d^4$ and $d^5$. 
Here, the XMCD spectrum has been obtained as the difference of the XAS spectra for the magnetic fields applied parallel and antiparallel to the photon helicity of the incident light, which is the equivalent to the difference between the spectra recorded for right-handed and left-handed circularly polarized light ($\mu_+ - \mu_-$) in a fixed magnetic field. 
Peak positions in the XMCD spectrum coincided with those of the multiplet structures of the XAS spectra. The XMCD spectra are also compared with the CI cluster-model calculations assuming the $d^5$ ground state and were well reploduced using the same parameter values. This suggests that the Mn$^{2+}$ ions in the Ga$_{1-x}$Mn$_x$N layers of the DFH sample as seen in the XAS spectra are responsible for the observed XMCD spectrum.

\begin{figure}[tttt]
\begin{center}
\includegraphics[height=4.5cm]{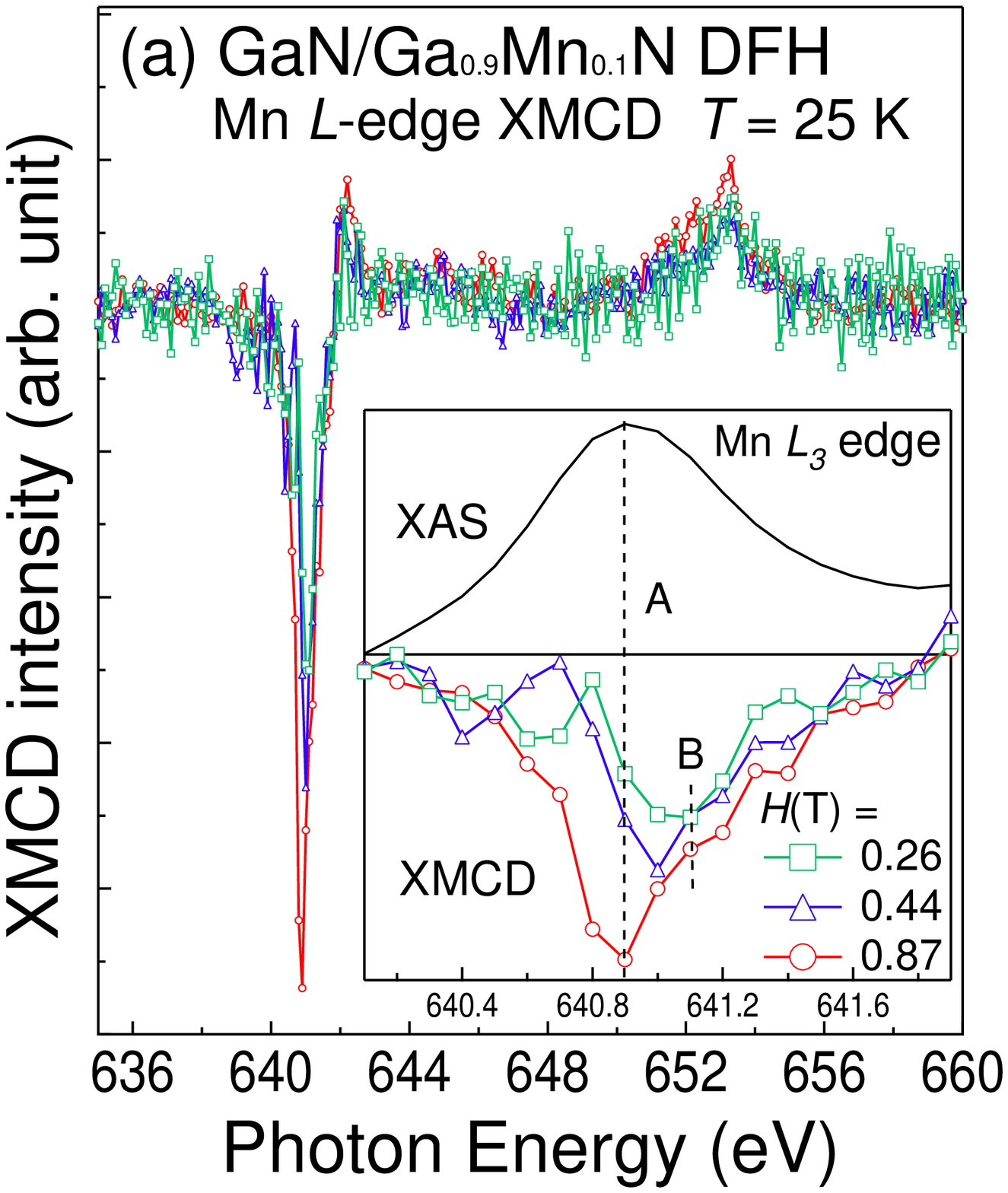}
\includegraphics[height=4.5cm]{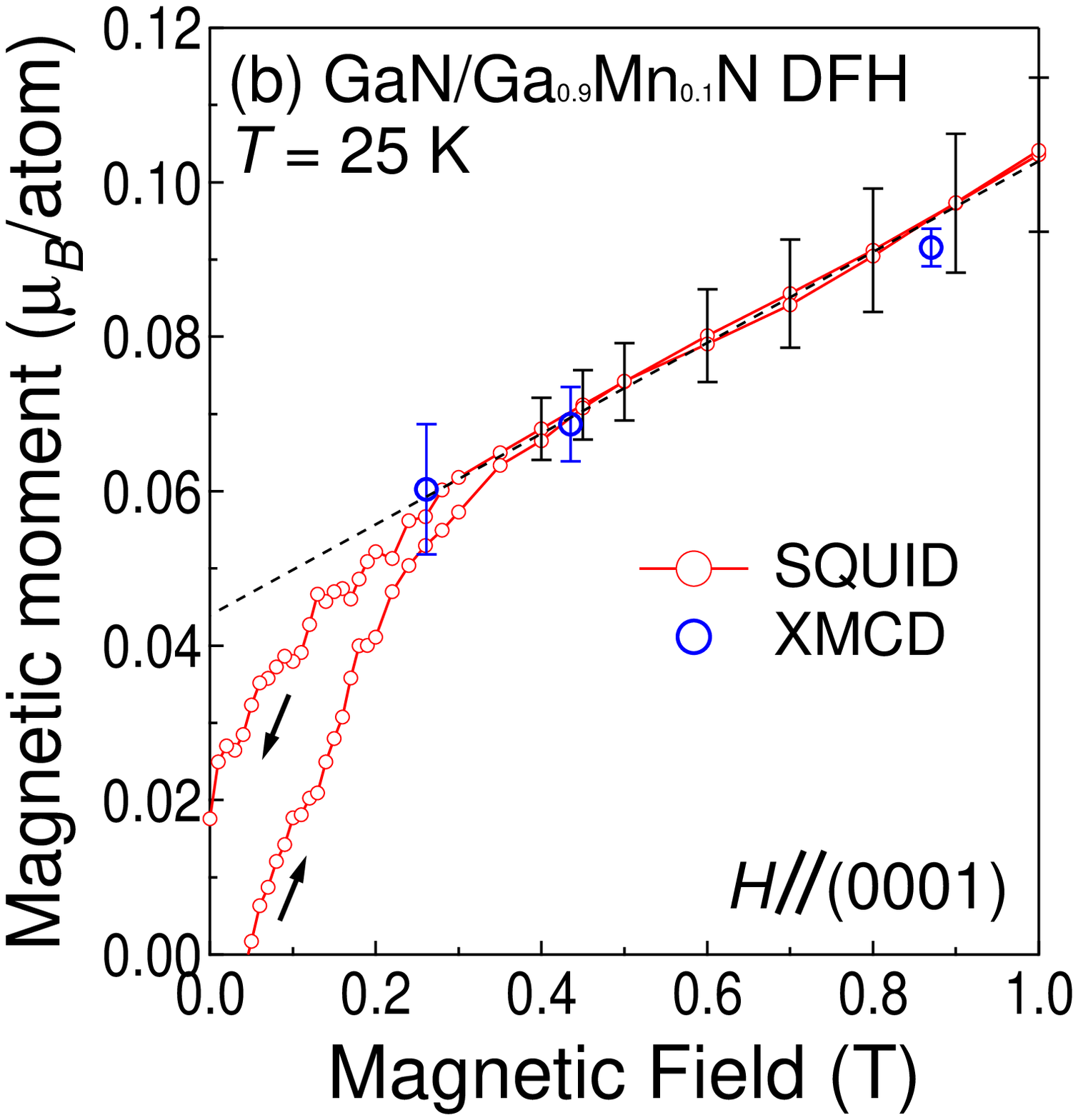}
\caption{(Color online) Magnetic field dependence of XMCD and the magnetization of GaN/Ga$_{1-x}$Mn$_x$N DFH at $T$ = 25 K. (a) Magnetic field dependence of the XMCD intensity. The inset shows an enlarged plot in the Mn $L_3$-edge region. (b) Magnetic moments per Mn atom obtained from the XMCD and SQUID data. The magnetic moment of $\sim$ 0.04 $\mu_B$ extrapolated to $H$ = 0 T corresponds to that of the SQUID data.}
\label{moment}
\end{center}
\end{figure}

To gain further information, the magnetic field dependence of XMCD was measured as shown in Fig. \ref{moment} (a). The intensities of the XMCD spectra decreased with decreasing magnetic field but no clear change in the overall spectral line shape was observed between the spectra recorded in different magnetic fields, as shown in the main panel of Fig. \ref{moment} (a).

Using the XMCD sum rules \cite{4sum1, 4sum2}, one can estimate the magnetic moment per Mn atom using the XAS and XMCD spectra. The magnetic moments at various magnetic fields thus obtained are plotted in Fig. \ref{moment} (b). Here, the XMCD intensities have been corrected for the degree of the circular polarization of 83 \%. The magnetic moment obtained using the SQUID magnetometer are also shown in Fig. \ref{moment} (b) 
Note that 
the diamagnetic contributions have been subtracted using the $\chi_0$ obtained from the Curie-Weiss fit and that the error bars of the SQUID data in Fig. \ref{moment} (b) are due to errors in $\chi_0$ [see Fig. \ref{squid} (d)]. In Fig. \ref{moment} (b), one can see excellent agreement between the SQUID and the XMCD data. 
The extrapolated magnetic moment of $M \sim 0.04\ \mu_B$/atom at $H$ = 0 was originated from the ferromagnetic component and the linear contribution from the paramagnetic component. Although the line shape of the XMCD spectra recorded at various magnetic fields were very similar, a subtle change was found in the $L_3$ main peak region as shown the inset of Fig. \ref{moment} (a). The XMCD intensity of the peak at $h\nu$ = 640.9 eV, which corresponds to the peak position of the Mn $L_3$ XAS spectra (labeled A) decreased with decreasing magnetic field while that at $h\nu$ = 641.2 eV (labeled B) was nearly independent of the magnetic field. This implies that in this magnetic field region the magnetic-field dependent (paramagnetic) and independent (ferromagnetic) components with slightly different electronic environments (of possibly caused by different local carrier concentration or defect concentrations) coexist and their electronic states are slightly different from each other. Also, one can conclude that the magnetically inactive component and the field-dependent component have similar electronic environment. 


In conclusion, we have performed XAS and XMCD measurements of a GaN/Ga$_{1-x}$Mn$_x$N DFH sample. From the XAS measurements, Mn ions in the Ga$_{1-x}$Mn$_x$N layer of DFH was found to be in the Mn$^{2+}$ state in a tetrahedral crystal field. From the XMCD measurements, the Mn ions were found to be responsible for the ferromagnetism of the sample. Comparison between the XMCD result and the SQUID data, it is suggested that the ferromagnetic as well as paramagnetic contributions arise from the Ga$_{1-x}$Mn$_x$N layer of this sample and cannot be attributed to any ferromagnetic precipitates.

This work was supported by a Grant-in-Aid for Scientific Research in Priority Area ``Semiconductor Nano-Spintronics'' (14076209). 
We also thank the Material Design and Characterization Laboratory,  Institute for Solid State Physics, University of Tokyo, for the use of the SQUID magnetometer.

\newpage

\vspace{1cm}

\vspace{1cm}

\end{document}